\documentclass[%
aps,
prl,
reprint,
superscriptaddress,
amsmath,amssymb,
longbibliography
]{revtex4-1}

\usepackage[pdftex]{graphicx}
\usepackage{color}
\usepackage{bm}
\usepackage[english]{babel}

\newcommand{\VWVE}{{VWVE}}
\newcommand{\minLCurve}{\tilde{\lambda}}
\newcommand{\dStot}{\Delta S_{\mathrm{tot}}}
\newcommand{\dSsys}{\Delta S_{\mathrm{sys}}}
\newcommand{\dSint}{\Delta S_{\mathrm{int}}}
\newcommand{\effCarnot}{\eta_{\mathrm{C}}}

\newcommand{\ReA}[1]{{\color{black} #1}}
\newcommand{\ReB}[1]{{\color{black} #1}}

\begin{document}

\title{Efficiency fluctuations in microscopic machines}

\author{Sreekanth K Manikandan}
\affiliation{Department of Physics, Stockholm University,\\SE-10691 Stockholm, Sweden.}

\author{Lennart Dabelow}%
\affiliation{Fakult\"at f\"ur Physik, Universit\"at Bielefeld, 33615 Bielefeld, Germany}

\author{Ralf Eichhorn}%
\affiliation{Nordita, Royal Institute of Technology and Stockholm University,
Roslagstullsbacken 23, SE-106 91 Stockholm, Sweden}

\author{Supriya Krishnamurthy}%
\affiliation{Department of Physics, Stockholm University,\\SE-10691 Stockholm, Sweden.}

\date{\today}

\begin{abstract}

Nanoscale machines are strongly influenced by thermal fluctuations, contrary to their macroscopic counterparts.
As a consequence, even the efficiency of such microscopic machines becomes a fluctuating random variable.
Using geometric properties and the fluctuation theorem for the total entropy production, a ``universal theory of efficiency fluctuations'' at long times, for machines with a finite state space, was developed in [Verley \textit{et al.}, Nat.~Commun.~\textbf{5}, 4721 (2014); Phys.~Rev.~E~\textbf{90}, 052145 (2014)].
We extend this theory to machines with an arbitrary state space. Thereby, we work out more detailed prerequisites for the ``universal features'' and explain under which circumstances deviations can occur. We also illustrate our findings with exact results for two non-trivial models of colloidal engines.

\end{abstract}

\maketitle

Understanding the functioning of machines on the micro- or nanoscale is
of great interest because of
their role in biological systems
and their numerous technological applications
\cite{Howard:2001mmp,parrondo2002energetics,Benenti:2017fas,Astumian:2012mra,
Blickle:2012rms, 
Dinis:2016tme, 
Krishnamurthy:2016msh}. 
Their small size makes this task a challenge since thermal fluctuations
strongly affect their operation.
As a result, average values are no longer
sufficiently informative, and fluctuations in
heat, work, efficiency \emph{etc.}\
must be taken into account.
Stochastic thermodynamics \cite{Seifert:2012stf} provides a convenient framework
for analyzing such systems by extending the notions of classical
(ensemble-based) thermodynamics to individual realizations of a given process.

Consider first a macroscopic heat engine operating cyclically between
two reservoirs
at different temperatures
$T_1 > T_2$ and performing work against an external load force.
If $Q_1$ and $Q_2$ denote the average heat exchanged with the two reservoirs
and $-W$ the (average) performed work, the Second Law
implies that the \emph{efficiency},
\begin{equation}
\label{eq:eta}
\eta = -W / Q_1
\, ,
\end{equation}
is universally bounded from above by the
\emph{reversible} or \emph{Carnot efficiency} $\effCarnot = 1 - T_2 / T_1$.

The efficiency $\eta$ plays an equally pivotal role 
for microscopic machines; however, in these systems, due to thermal fluctuations,
the value obtained in
individual realizations 
can deviate significantly from the average behavior. We hence need to
consider a distribution of efficiency values.
Recently, in two seminal papers \cite{Verley:2014uce,Verley:2014ute},
Verley, Willaert, Van den Broeck, and Esposito (\VWVE) developed a
``universal theory of efficiency fluctuations'' for machines with a finite state space.
By characterizing the long-time behavior of the efficiency fluctuations
in terms of their large-deviation function $J(\eta)$
(see below for more details),
they found that the 
\ReA{\emph{macroscopic efficiency}, defined as the ratio of average output and input powers,}
is the most likely and,
for machines operating in a non-equilibrium steady state or under a time-symmetric periodic protocol,
the reversible Carnot efficiency is the least likely one
\footnote{For machines driven asymmetrically in time, the reversible efficiency stands out as the value of $\eta$ at which the rate functions $J(\eta)$ and $\tilde J(\eta)$ for the forward and time-reversed drivings, respectively, intersect.}.
The \VWVE\ theory has since been verified in numerous model systems with finite
\cite{Gingrich:2014eld,
Proesmans:2015sef,
Polettini:2015esa,
Proesmans:2015esp,
Esposito:2015efq,
Cuetara:2015dqd,
Jiang:2015esb,
Agarwalla:2015fcs%
}
but also infinite
\cite{Proesmans:2015sef,
Proesmans:2015see,
Agarwalla:2015fcs,
sune2018efficiency%
}
state spaces.

Nevertheless, there are a few examples of infinite state space systems at odds with the theory
\cite{Park:2016emp,
Gupta:2018edw,
Gupta:2017sei,
vroy:tel-01968075},
in which the rate function $J(\eta)$ fails to be smooth and/or does not exhibit a unique maximum
at the reversible efficiency. A clear understanding of why some systems with infinite state space
obey the ``universal'' theory while others do not is lacking. In this Letter, we give detailed prerequisites for when the features of the VWVE theory
are found and when they are violated. In doing so, we develop an extended general theory of efficiency fluctuations, unifying the VWVE theory with deviations observed in specific models. Two examples of analytically solvable machines \cite{Gupta:2017sei,Filliger:2007bgm}
serve as illustrations for our general findings.

We start by briefly summarizing the approach taken in the \VWVE\ theory. 
For all systems, the total work $W$ and heat $Q_{1}$
(as well as $Q_{2}$) grow extensively with increasing operational
time $\tau$. 
For microscopic systems they also naturally fluctuate
due to thermal noise, leading to a distribution $p_\tau(q_1,w)$
for observing a heat absorption rate $q_1=Q_1/\tau$ (the input power)
and an output power $-w=-W/\tau$ 
 with average values
 $\langle q_{1} \rangle$ and $\langle -w \rangle$.
Using the theory of large deviations \cite{Touchette:2009lda},
we can quantify the asymptotic decay of the probability $p_\tau(q_1,w)$
towards the delta-distribution peaked at $\langle q_1 \rangle$ and $\langle w \rangle$ 
by the
\emph{large deviation} or \emph{rate function} $I(q_1, w)$,
\begin{equation}
\label{eq:RateFuncQW}
	p_\tau(q_1, w) \sim e^{-\tau \, I(q_1, w)} \quad (\tau \to \infty)
\, ,
\end{equation}
where $I(q_1, w) \geq 0$ and $I(\langle q_1 \rangle, \langle w \rangle) = 0$.
Similarly, the stochastic efficiency $\eta = -w / q_1$ will tend towards
the
\ReA{macroscopic efficiency}
$\bar\eta = \langle -w \rangle / \langle q_1 \rangle$.
Again, we can describe this approach in terms of a rate function $J(\eta)$,
providing an asymptotic relation for the probability distribution $p_\tau(\eta)$,
\begin{equation}
\label{eq:RateFuncEta}
	p_\tau(\eta) \sim e^{-\tau J(\eta)} \quad (\tau \to \infty)
\, .
\end{equation}
This $J(\eta)$ can be extracted from the
scaled cumulant generating function (sCGF) of heat and work \cite{Verley:2014uce,Verley:2014ute},
\begin{equation}
\label{eq:DefPhi}
	\phi(\lambda_Q, \lambda_W) :=
	\lim_{\tau\to\infty} \frac{1}{\tau} \ln \left\langle e^{-\lambda_Q Q_1 - \lambda_W W} \right\rangle_\tau
\, ,
\end{equation}
according to
\begin{equation}
\label{eq:JFromPhi}
	J(\eta) = - \min_{\lambda} \phi(\eta \lambda, \lambda)
\, .
\end{equation}
Here $\langle \cdots \rangle_\tau$
denotes an average over 
the distribution $p_\tau(q_1, w)$. Note that $\phi(\lambda_Q, \lambda_W)$ is a
convex function by definition.

\begin{figure}
\includegraphics[scale=1]{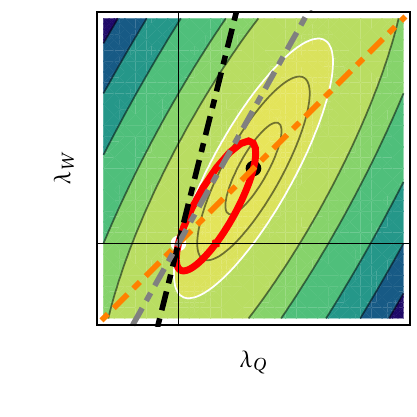}
\includegraphics[scale=1]{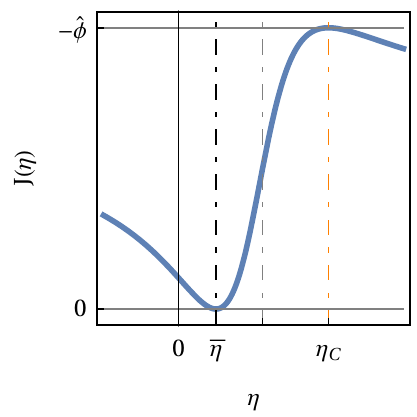}
\caption{Illustration of the relation~\eqref{eq:JFromPhi} between $\phi(\lambda_Q, \lambda_W)$ and $J(\eta)$.
Left: Contour plot of $\phi(\lambda_Q, \lambda_W)$ showing its convexity along
with the lines $\lambda_Q = \eta \lambda_W$ (dashed) for the
macroscopic efficiency $\bar\eta$ (black), reversible efficiency $\effCarnot$ (orange)
and an intermediate value (gray).
The red curve marks the minimizing $\bm{\minLCurve}(\eta)$ for all $\eta$.
Right: The resulting $J(\eta) = -\phi(\minLCurve_Q(\eta), \minLCurve_W(\eta))$.
}
\label{fig:JFromPhi}
\end{figure}
The relation \eqref{eq:JFromPhi} implies $\phi(0, 0) = 0 \leq J(\eta) \leq -\hat\phi$,
where $\hat\phi := \min_{\lambda_Q, \lambda_W} \phi(\lambda_Q, \lambda_W)$
is the global minimum of $\phi$. 
Moreover, it has the geometric interpretation, illustrated in Fig.~\ref{fig:JFromPhi}:
For fixed $\eta$, We obtain $J(\eta)$ by minimizing
$\phi(\lambda_Q, \lambda_W)$ along the line $\lambda_Q = \eta \lambda_W$ and inverting the sign.
The set of all points $\bm\minLCurve(\eta) \equiv (\minLCurve_Q(\eta), \minLCurve_W(\eta))$
where the minima are attained as a function of $\eta$
describes a curve in the $(\lambda_Q,\lambda_W)$-plane
(see Fig.~\ref{fig:JFromPhi})
with $J(\eta) = -\phi(\minLCurve_Q(\eta), \minLCurve_W(\eta))$.

Exploiting this geometrical picture,
the aforementioned ``universal theory'' by \VWVE~\cite{Verley:2014uce,Verley:2014ute}
establishes generic properties of $J(\eta)$
that are independent of system-specific details.
As the main result they find that
$J(\eta)$ is a smooth function with a unique minimum
at the macroscopic efficiency $\bar\eta$, such that $J(\bar\eta) = 0$, and a unique maximum at some finite efficiency $\hat\eta$.
For time-symmetric driving protocols,
this ``least likely'' efficiency $\hat\eta$ coincides with the reversible efficiency $\effCarnot$
\cite{Note1},
see the example in Fig.~\ref{fig:JFromPhi}.

These results of the \VWVE\ theory are based on a few
assumptions, most notably:
(i) The detailed fluctuation theorem \cite{Seifert:2005epa} $p(\dStot)/p(-\dStot) = \exp(\dStot)$ for the
total entropy production  $\dStot= -Q_1/T_1 - Q_2/T_2 + \dSsys$
(where $\dSsys$ denotes the entropy change in the system itself) is valid,
(ii) $\phi(\lambda_Q, \lambda_W)$ is a smooth function of its arguments
and the fluctuation theorem implies that it
has the symmetry property
\begin{equation}
\label{eq:PhiSymmetry}
	\phi(\lambda_Q, \lambda_W) = \phi(\lambda_Q^* - \lambda_Q, \lambda_W^* - \lambda_W)
\end{equation}
with $\lambda_Q^* = \effCarnot / T_2$ and $\lambda_W^* = 1/T_2$, and
(iii) the minimum of $\phi(\lambda_Q, \lambda_W)$ is unique.
The validity of (i) is by now well-established 
\cite{Seifert:2005epa,Seifert:2012stf}. 
However, we will demonstrate that it does not necessarily entail
the validity of the symmetry \eqref{eq:PhiSymmetry} for all $(\lambda_Q,\lambda_W)$
as in (ii).
Further, we discuss the case that assumption (iii) does not hold either.

While assumption (ii) appears plausible for systems with finite state space,
it has been observed in certain models with infinite state space that the sCGF
\eqref{eq:DefPhi} can have a restricted domain of convergence $C_0$ \cite{Park:2016emp,Gupta:2017sei}. It has also been noticed that the symmetry property \eqref{eq:PhiSymmetry} need not necessarily hold \cite{Gupta:2017sei,vanzone2,jdnoh,ever,sabha1,sbpflc}.
To clarify the relationship between a limited convergence domain and the symmetry relation~\eqref{eq:PhiSymmetry}, we
express the sCGF in terms of the individual time-extensive and -intensive contributions
to the total entropy production,
\begin{equation}
\label{eq:StotSplit}
	\dStot = \frac{\effCarnot}{T_2} Q_1 + \frac{1}{T_2} W + \dSint
\, .
\end{equation}
Here, the term $\dSint = -\frac{1}{T_2} \Delta E + \dSsys$
collects the time-intensive contributions to the total entropy production that depend only on the initial and final states of the system.
$\Delta E$ denotes the change in internal energy, which is,
according to the First Law, $\Delta E = W + Q_1 + Q_2$.

We first write down the moment-generating function (MGF)
for the combined probability distribution of
the individual contributions from \eqref{eq:StotSplit},
\begin{equation}
\label{eq:DefMGF}
	\Psi_\tau(\lambda_Q, \lambda_W, \lambda_S) := \left\langle e^{-\lambda_Q \, Q_1 - \lambda_W \, W - \lambda_S \, \dSint} \right\rangle_\tau
\, .
\end{equation}
The fluctuation theorem for the total entropy production implies that
$\Psi_\tau$
has the symmetry property \cite{Lebowitz:1999gct,Garcia-Garcia:2010uaf}, $\Psi_\tau(\lambda_Q, \lambda_W, \lambda_S) = \Psi_\tau(\lambda_Q^* - \lambda_Q, \lambda_W^* - \lambda_W, 1 - \lambda_S)$.
This symmetry is inherited by the three-dimensional sCGF
\begin{equation}
\label{eq:DefTotalsCGF}
	\phi(\lambda_Q, \lambda_W, \lambda_S) := \lim_{\tau\to\infty} \frac{1}{\tau} \ln \Psi_\tau(\lambda_Q, \lambda_W, \lambda_S)
\, ,
\end{equation}
so that
\begin{equation}
\label{eq:TotalsCGFSymmetry}
	\phi(\lambda_Q, \lambda_W, \lambda_S) = \phi(\lambda_Q^* - \lambda_Q, \lambda_W^* - \lambda_W, 1 - \lambda_S)
\, .
\end{equation}
The sCGF \eqref{eq:DefPhi} of heat and work alone is the restriction of that
``total'' sCGF to the $\lambda_S = 0$ plane, \emph{i.e.}\
$\phi(\lambda_Q, \lambda_W) = \phi(\lambda_Q, \lambda_W, 0)$.

As a consequence of Eq. \eqref{eq:TotalsCGFSymmetry} we arrive at 
the central observation that this restricted sCGF fulfills a ``symmetry'' relation of the form
$\phi(\lambda_Q,\lambda_W)=\phi(\lambda_Q,\lambda_W,0)=\phi(\lambda_Q^* - \lambda_Q, \lambda_W^* - \lambda_W, 1)$, instead of the relation \eqref{eq:PhiSymmetry}.  However, \eqref{eq:PhiSymmetry} could still be valid if
$\phi(\lambda_Q,\lambda_W,\lambda_S)$ were independent of $\lambda_S$.
Indeed, the \VWVE\ theory \cite{Verley:2014uce,Verley:2014ute} is restricted to
machines with a \emph{finite state space},
in which case both, the internal energy $\Delta E$ and the system entropy $\dSsys$, are bounded
by $\tau$-independent constants, implying that the $\lambda_S$ dependence in
\eqref{eq:DefTotalsCGF} disappears in the $\tau\to\infty$ limit.

\begin{figure}
\centering
\includegraphics[scale=0.3]{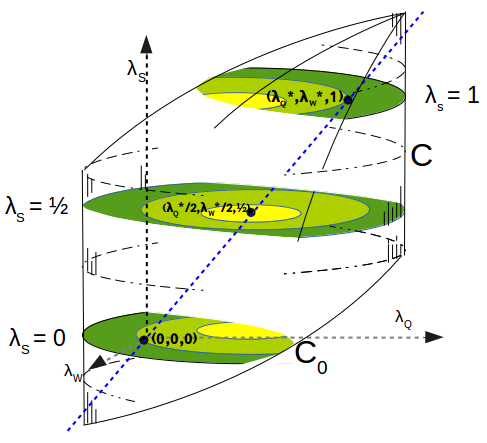}
\caption{Contour plots of a typical $\phi(\lambda_Q, \lambda_W, \lambda_S)$
for $\lambda_S = 0, \frac{1}{2}, 1$ along with the domain of convergence $C$ and the domain of convergence $C_0$ for $\phi$ at $\lambda_S=0$.
The functional form of $\phi(\lambda_Q, \lambda_W, \lambda_S)$ is the same in
all $\lambda_S = \mathrm{const}$ planes, but the limited domain of convergence
leads to cutoffs whose location changes as a function of $\lambda_S$.
The symmetry around the point $(\lambda_Q^*/2,\lambda_W^*/2,1/2)$
is a consequence of the fluctuation theorem.
\label{fig:TotalsCGF}
}
\end{figure}

However, if fluctuations of
the intensive entropy production $\dSint$ can become arbitrarily large,
as is typically the case for machines with infinite state space
\cite{Gupta:2017sei,Zon:2004ehf,Noh:2014fcn,Lee:2013eim,Verley:2014wss,Sabhapandit:2011wfh,Pal:2013wfb},
we cannot argue that $\phi$ is independent of $\lambda_S$.
In this case, too, 
the $\lambda_S$ contributions drop out in Eq.~\eqref{eq:DefTotalsCGF} as $\tau\to\infty$
wherever $\Psi_\tau(\lambda_Q,\lambda_W,\lambda_S)$ remains real and finite.
However, in contrast to the finite state-space case,
the limited domain of convergence $C$ of $\Psi_\tau$
will in general depend on $\lambda_S$.
As a consequence of the fluctuation theorem symmetry obeyed by $\Psi_{\tau}$,
$C$ is symmetric about the point $(\lambda_Q^*/2, \lambda_W^*/2, 1/2)$.
Therefore, the restriction of $C$ to the $\lambda_S = 1/2$
plane satisfies a symmetry property like~\eqref{eq:PhiSymmetry},
but the domain of convergence $C_0$ of $\phi$ at $\lambda_S=0$
will in general not obey this symmetry,
and hence neither will $\phi(\lambda_Q, \lambda_W, 0)$.
This situation is illustrated in Fig.~\ref{fig:TotalsCGF}.



%

What are the consequences of the limited domain of convergence $C_0$ and its
lack of symmetry for the large deviation function $J(\eta)$?
The answer depends on whether the minimizing curve $\bm\minLCurve(\eta)$ is completely
contained inside $C_0$ or whether it touches or hits the boundary of $C_0$.
We illustrate this difference using the
isothermal work-to-work converter from Ref.~\cite{Gupta:2017sei},
\ReA{which exhibits both of these cases depending on the amplitude ratio of work and load forces}.
This machine
consists of a Brownian particle in contact with a single heat bath at temperature $T$
and two additional white-noise forces, interpreted as a load and drive force, respectively.
Identifying the work done by the drive force with $Q_1$ 
and the work done by the load force with $W$, we can calculate the sCGF $\phi(\lambda_Q, \lambda_W)$ exactly, and from that the curve $\bm\minLCurve(\eta)$ and rate function $J(\eta)$ (see \cite{Gupta:2017sei} and the Supplemental Material~\cite{suppl} \textcolor{black}{which includes references~\cite{Gartner:1977old,Ellis:1984ldf}} for details).

\begin{figure}
\centering
\includegraphics[scale=1]{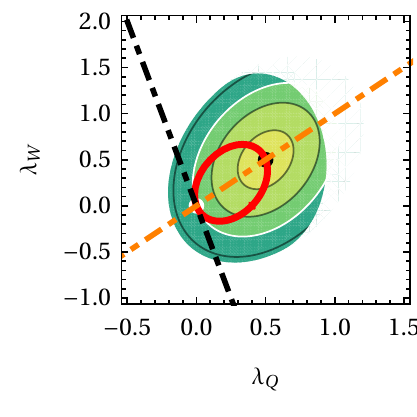}
\includegraphics[scale=1]{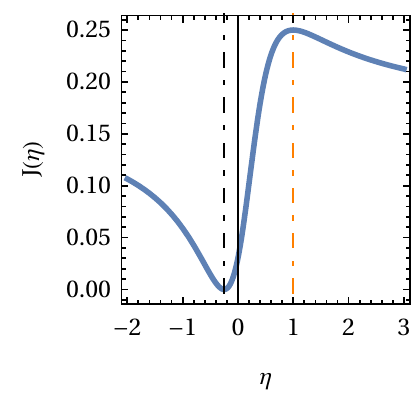} \\
\includegraphics[scale=1]{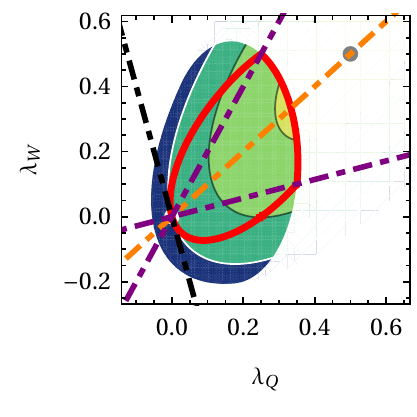}
\includegraphics[scale=1]{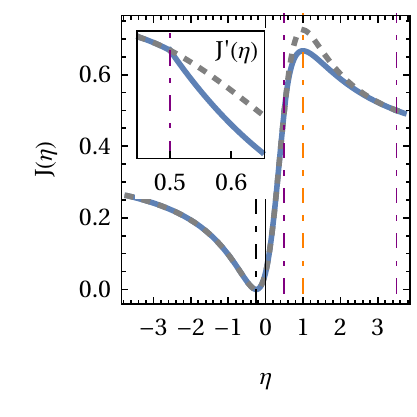}
\caption{$\phi(\lambda_Q, \lambda_W)$ (left) and $J(\eta)$ (right) for the
isothermal work-to-work converter from Ref.~\cite{Gupta:2017sei}. Exact results are shown for two different parameter sets of the model
(see~\cite{Gupta:2017sei, suppl} for details).
Dashed black lines mark the corresponding macroscopic efficiencies $\bar\eta$
(negative in this case%
),
dashed orange lines the reversible efficiency $\effCarnot=1$.
Top (model parameters $\theta=1$, $\alpha=1/2$): 
Since the cutoff does not intersect the minimizing curve $\bm\minLCurve(\eta)$,
the emerging shape of the rate function $J(\eta)$ still follows the predictions
of the \VWVE\ theory: It is smooth and has a unique maximum at the reversible efficiency $\effCarnot = 1$.
Bottom ($\theta=4$, $\alpha=1/2$):
The cutoff interferes with the minimizing curve $\bm\minLCurve(\eta)$
so that the latter runs along the boundary of the domain of convergence
for \ReB{$\eta \in [\frac{1}{2}, \frac{7}{2}]$ (purple, dashed lines)} and the rate function (right panel, blue curve)
becomes distorted from the shape that would be obtained without cutoffs (dotted, gray curve).
In particular, it develops kinks that become visible in its first derivative. \ReB{One of these} is displayed in the inset of the bottom-right panel.
} 
\label{fig:W2WConv}
\end{figure}

In the first case, when $\bm\minLCurve(\eta)$ lies completely inside $C_0$,
the existence of singular points for $\phi(\lambda_Q, \lambda_W)$ is irrelevant,
resulting in a $J(\eta)$ that has exactly the properties and ``universal shape''
predicted by the \VWVE\ theory (see the top panels in Fig.~\ref{fig:W2WConv}).
In particular, the reversible efficiency $\effCarnot = 1$
is still least likely,
because the global minimum $\hat\phi$ of $\phi$ is still attained
at the point $(\lambda_Q^*/2,\lambda_W^*/2)$
despite the ``asymmetry'' of $C_0$.
By contrast, in the second case,
$\phi$ takes its minimal value on the boundary of $C_0$ (lower left panel in Fig.~\ref{fig:W2WConv}).
The minimizing curve $\bm\minLCurve(\eta)$ thus follows the boundary of $C_0$
for some range of $\eta$ values and becomes non-smooth at the points where
\ReB{the path transitions from the interior to the boundary and vice versa},
leading to kinks in
\ReB{the first derivative of $J(\eta)$}
(lower right panel in Fig.~\ref{fig:W2WConv}).
We conclude that the appearance of cutoffs in $\phi(\lambda_Q, \lambda_W)$
can lead to discontinuities or ``kinks'' in $J(\eta)$ or its derivatives.
In general (see also \cite{suppl}), $J(\eta)$ is a smooth function of $\eta$ if and only if
$\phi(\lambda_Q, \lambda_W)$ is smooth along the curve $\bm\minLCurve(\eta)$.
Note that in the second example of Fig.~\ref{fig:W2WConv} (lower panels), 
the
least likely efficiency is still $\effCarnot$, even though $\phi(\lambda_Q,\lambda_W)$ does
not obey the symmetry \eqref{eq:PhiSymmetry}.
However, this need not be the case in general since the minimal $\phi$-value need no longer be located on the line $\lambda_Q=\effCarnot\lambda_W$. 
%

\begin{figure}
\centering
\includegraphics[scale=1]{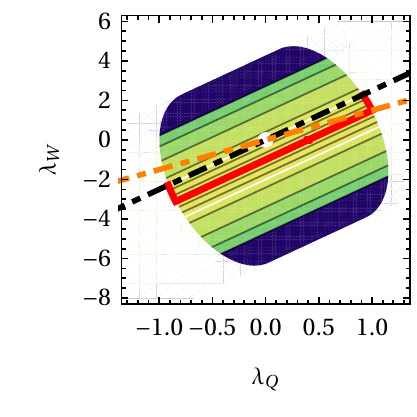}
\includegraphics[scale=1]{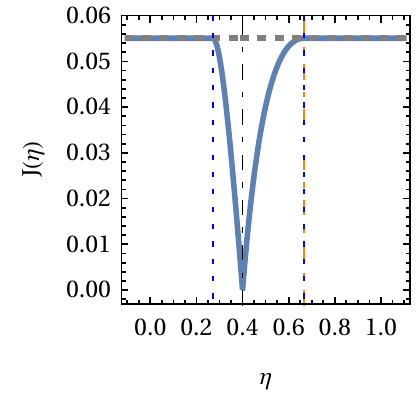}
\\
\includegraphics[scale=1]{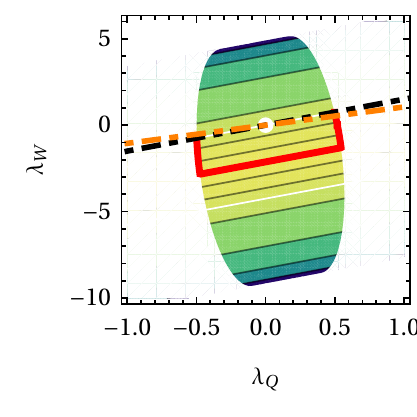}
\includegraphics[scale=1]{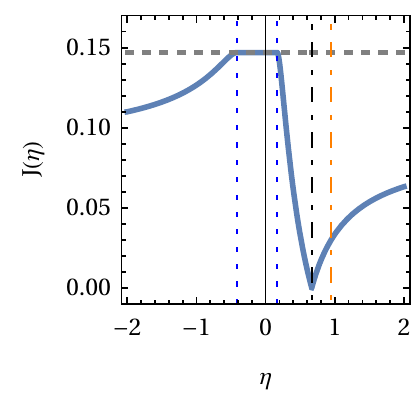}
\caption{$\phi(\lambda_Q,\lambda_W)$ (left) and $J(\eta)$ (right) for two configurations of the Brownian gyrator.
Dashed black lines mark the macroscopic efficiency $\bar\eta$,
dashed orange lines the reversible efficiency $\effCarnot$.
The minimizing curve $\bm\minLCurve(\eta)$ is shown in red in the left panels.
The efficiencies at the edges of the plateau region are marked by dotted blue lines in the right panels.
Top: $u_1 = 5$, $u_2 = 1$, $\alpha = \pi/4$, $f_{\mathrm{ext}} = -1/2$, $k_{\mathrm B} T_1 = 1$, $k_{\mathrm B} T_2 = 1/3$.
Bottom: $u_1 = 4$, $u_2 = 2$, $\alpha = \pi/4$, $f_{\mathrm{ext}} = -1/2$, $k_{\mathrm B} T_1 = 2$, $k_{\mathrm B} T_2 = 1/10$.
In all plots, $\gamma_1 = \gamma_2 = 1$. \ReB{In this case, the kinks occur in $J(\eta)$ itself.}} 
\label{fig:Gyrator}
\end{figure}

Next, we investigate the situation when assumption (iii) fails to hold
and the global minimum
of $\phi(\lambda_Q, \lambda_W)$
is not unique but rather degenerate
\footnote[2]{Degeneracy in $\phi$ in systems with both tight coupling and dynamical phase transitions has been investigated in \cite{vroy:tel-01968075}},
\emph{i.e.}\
there exist multiple points $(\lambda_Q, \lambda_W)$ in the set
$\hat R:=\{(\lambda_Q, \lambda_W)|\phi(\lambda_Q, \lambda_W)=\hat\phi\}$.
Due to the convexity of $\phi(\lambda_Q, \lambda_W)$,
this set will be a connected region in the $(\lambda_Q, \lambda_W)$-plane.
Then $J(\eta)$ assumes its maximal value $-\hat\phi$ for all $\eta$
for which the line $\lambda_Q = \eta\lambda_W$ intersects the region $\hat R$,
leading to a plateau of degenerate maxima.
\ReB{Presumably, such a scenario could also occur in systems with finite state space.}
The reversible efficiency $\effCarnot$ is one of these maximizing efficiencies
if and only if the line $\lambda_Q = \effCarnot \lambda_W$
intersects the region $\hat R$ within the domain of convergence $C_0$ (see also \cite{suppl}).

We illustrate this situation with the example of the ``Brownian gyrator''
\cite{Filliger:2007bgm,Pietzonka:2018utp}. 
This heat engine consists of a colloidal particle in two dimensions,
immersed in a fluid environment and experiencing thermal fluctuations
of different intensity along two perpendicular directions
(temperatures $T_1$ and $T_2$, friction coefficients $\gamma_1$ and $\gamma_2$;
see~\cite{Argun:2017erm,Chiang:2017eab} for experimental realizations).
The particle is trapped in a harmonic potential whose principal axes
with stiffnesses $u_1$ and $u_2$ are rotated by an angle $\alpha$
with respect to the preferred axes of the heat baths.
As a consequence, the particle experiences a net torque 
letting it rotate around the origin on average \cite{Filliger:2007bgm}.
Applying a linear ``load torque'' with slope $f_{\mathrm{ext}}$,
the system operates as a stationary, heat engine \cite{Pietzonka:2018utp}
(see also \cite{suppl} for details).
The resulting sCGF $\phi(\lambda_Q, \lambda_W)$ and rate function $J(\eta)$ can be computed exactly using path-integral techniques~\cite{Manikandan:2017awd,Manikandan:2018erf} \textcolor{black}{( see also \cite{suppl}, which includes references ~\cite{Onsager:1953fai,Machlup:1953fai,Chernyak:2006pia,Kirsten:2003fdc} )}
and are shown for two different configurations in Fig.~\ref{fig:Gyrator}.
In this system, the degenerate minimum of $\phi(\lambda_Q, \lambda_W)$
results from $\phi(\lambda_Q, \lambda_W)$
becoming a function of only $\lambda_Q - \bar\eta \lambda_W$
within the domain of convergence $C_0$
due to  tight
coupling between work and heat \cite{Polettini:2015esa}. 
The iso-contours of $\phi$, one of them being the set $\hat R$, are therefore parallel lines with slope $1/\bar\eta$ (see left panels of Fig.~\ref{fig:Gyrator}).
The resulting plateaus for $J(\eta)$ are  visible in the right panels.
For the configuration in the top panels, the region $\hat R$ intersects the
$\lambda_Q$-axis, so that the plateau of $J(\eta)$ extends to $\pm\infty$.
Moreover, in this configuration the line $\lambda_Q = \effCarnot \lambda_W$
intersects $\hat R$, so that the Carnot efficiency lies at the edge of
the plateau of degenerate maxima of $J(\eta)$. 
In contrast, for the second configuration in the lower panels,
$\hat R$ does not intersect the $\lambda_Q$-axis and the plateau is restricted
to a finite region of $\eta$ values.
Furthermore, this plateau does not contain the Carnot efficiency $\effCarnot$.
We note that in both cases $J(\eta)$ has kinks resulting from the minimizing
curve $\bm\minLCurve(\eta)$ hitting the boundary of the domain of convergence $C_0$. A similar efficiency distribution has also been obtained in \cite{Park:2016emp} for a closely related model.

In conclusion, we have extended the VWVE theory of
efficiency fluctuations by including three crucial insights: First,
the domain of convergence $C_0$ of $\phi(\lambda_Q, \lambda_W)$ determines if
the fluctuation theorem symmetry \eqref{eq:PhiSymmetry} holds or not.
The resulting cutoffs lead to a $J(\eta)$ differing from the VWVE theory,
if and only if $\phi$ is non-smooth along the curve $\bm\minLCurve(\eta)$.
\ReB{Since $C_0$ depends on the boundary terms, this opens up
possibilities to fine-tune initial conditions to change $J(\eta)$
and, for example, minimize fluctuations around the macroscopic efficiency.}
Secondly, $\phi$ can have degenerate minima, typically leading to a
plateau of maximal values for $J(\eta)$.\ReB{Finally, the symmetry
\eqref{eq:PhiSymmetry} is sufficient, but not necessary for the
least likeliness of the Carnot efficiency $\effCarnot$.
On the other hand, $\effCarnot$ is no longer least likely in general.} Our results are obtained from generic features of the sCGF $\phi$ and incorporate as special cases the VWVE theory as well as exceptions found in specific models. Exact solutions for two non-trivial models support our observations.

\begin{acknowledgments}
\emph{Acknowledgements.}
LD acknowledges funding by the Deutsche Forschungsgemeinschaft within the
Research Unit FOR~2692 under Grant No. 397303734 as well as by the
Stiftung der Deutschen Wirtschaft.
LD also thanks Nordita for the hospitality and support during an internship in 2016.
RE acknowledges financial support from the Swedish Research Council
(Vetenskapsr{\aa}det) under the grant No.~2016-05412.
\end{acknowledgments}

\begin{widetext}
\section*{Supplemental material}
This document provides further details of the calculations behind the results presented in the manuscript 
``Efficiency fluctuations in microscopic machines''.
In the first section, we further elaborate the relationship between properties of the 
scaled cumulant generating function of heat and work $\phi(\lambda_Q, \lambda_W)$
and properties of the efficiency rate function $J(\eta)$.
In the second section, we summarize the essential findings (relevant for our analysis)
of Ref.~\cite{Gupta:2017sei} about the isothermal work-to-work converter used as an
illustrative example in the main text.
In the third section, we introduce the Brownian gyrator model, which served as a second
illustrative example in the main text, and present details on the derivation of its scaled
cumulant generating function of heat and work.

\section*{Relation between properties of $\phi(\lambda_Q, \lambda_W)$ and properties of $J(\eta)$}
In this section, we formalize how certain prerequisites for the scaled
cumulant generating function (sCGF) of heat and work $\phi(\lambda_Q, \lambda_W)$
lead to properties of the efficiency rate function $J(\eta)$, namely, smoothness, the least likely efficiency and plateaus.

Before we turn to the specific observations from the main text,
we collect a few basic properties of the sCGF $\phi(\lambda_Q, \lambda_W)$.
By definition (see Eq. (4) in the main text ), $\phi(\lambda_Q, \lambda_W)$
is a convex function, notably meaning that the sublevel sets
\begin{equation}
\label{eq:PhiSublevelSets}
	A_r := \left\{ (\lambda_Q, \lambda_W) \in \mathbb R^2 : \phi(\lambda_Q, \lambda_W) \leq r \right\}
\end{equation}
are convex.
Moreover, it satisfies the normalization condition $\phi(0, 0) = 0$.
As $J(\eta)$ is obtained from $\phi(\lambda_Q, \lambda_W)$ by minimizing
along lines $\lambda_Q = \eta \lambda_W$ through the origin and inverting
the sign (see Eq. (5) in the main text), this immediately implies $J(\eta) \geq 0$.
It also implies that the curve $\bm\minLCurve(\eta)$ is contained in the sublevel set $A_0$.

Furthermore, for large deviation theory to be applicable to the problem at all,
we need that the rate function $I(q_1, w)$, \ReA{defined via Eq.~(2) from the main paper,}
is well-defined and, in particular, $I(0, 0)$ has a finite value.
This in turn implies, according to the G\"artner-Ellis theorem \cite{Gartner:1977old, Ellis:1984ldf, Touchette:2009lda}
$I(q_1, w) = \max_{\lambda_Q, \lambda_W} \left[ \lambda_Q \, q_1 + \lambda_W \, w - \phi(\lambda_Q, \lambda_W) \right]$,
that $\phi(\lambda_Q, \lambda_W)$ is bounded from below, so that there is a value
$\hat\phi \in \mathbb R$ with $\phi(\lambda_Q, \lambda_W) \geq \hat\phi$ for all $\lambda_Q$, $\lambda_W$.
All these properties will be taken for granted in the following.

\subsection*{Smoothness}
As argued in the main text, $J(\eta)$ is a smooth function of $\eta$ if and only
if $\phi(\lambda_Q, \lambda_W)$ is smooth along the curve $\bm\minLCurve(\eta)$.
This follows immediately from the definition of
\ReA{$J(\eta) = -\min_\lambda \phi(\eta\lambda, \lambda)$ [see also Eq.~(5) from the main paper]}
and
from the definition of $\bm\minLCurve(\eta)$,
implying $J(\eta) = -\phi(\minLCurve_Q(\eta), \minLCurve_W(\eta))$.
However, this criterion is not very ``practical'' because it generally becomes
quite complicated to determine the curve $\bm\minLCurve(\eta)$ in the presence
of singular points for $\phi(\lambda_Q, \lambda_W)$.
A more accessible (albeit weaker) characterization is as follows:

\textit{If the global minimum of $\phi(\lambda_Q, \lambda_W)$ is unique and there
exists an open region $U \subseteq \mathbb R^2$ with $A_0 \subseteq U$ such that
$\phi(\lambda_Q, \lambda_W)$ is smooth in $U$ and the Hessian matrix of $\phi(\lambda_Q, \lambda_W)$
is positive definite in $U$, then $J(\eta)$ is smooth.}

This provides a sufficient (but not necessary) condition on $\phi(\lambda_Q, \lambda_W)$
for $J(\eta)$ to be smooth.
To prove this, assume that $\phi(\lambda_Q, \lambda_W)$ is smooth in some open region $U$
containing the sublevel set $A_0$. Moreover, assume that the Hessian matrix of $\phi(\lambda_Q, \lambda_W)$
is positive definite on $U$, implying that the function is strictly convex on $U$.
Smoothness means that $\phi(\lambda_Q, \lambda_W)$ is infinitely differentiable for all $(\lambda_Q, \lambda_W) \in U$.
For every $\eta$, denote by
$\bm\minLCurve(\eta) \equiv (\minLCurve_Q(\eta),  \minLCurve_W(\eta))$ a point with
$\minLCurve_Q(\eta) = \eta \minLCurve_W(\eta)$ that minimizes Eq. (5), \emph{i.e.}\ $\phi(\minLCurve_Q(\eta), \minLCurve_W(\eta)) = \min_{\lambda} \phi(\eta\lambda, \lambda)$.
As observed above, $J(\eta) \geq 0$ for all $\eta$, so that $\bm\minLCurve(\eta) \in A_0 \subseteq U$.
Obviously, $J(\eta) = -\phi(\minLCurve_Q(\eta), \minLCurve_W(\eta))$, meaning that the function
$J(\eta)$ is determined by the values of $\phi(\lambda_Q, \lambda_W)$ on the curve
$\eta \mapsto \tilde{\bm\lambda}(\eta)$ with $\eta \in \mathbb R$.
It suffices to show that this mapping is smooth.
The smoothness of $\phi(\lambda_Q, \lambda_W)$ in $U$ then implies that
$J(\eta) = -\phi(\minLCurve_Q(\eta), \minLCurve_W(\eta))$ is smooth as well.

Smoothness and convexity of $\phi(\lambda_Q, \lambda_W)$ imply that the line
$\lambda_Q = \eta\lambda_W$ is tangent to the iso-contour
$\phi(\lambda_Q, \lambda_W) = J(\eta)$ in the point $\tilde{\bm\lambda}(\eta)$.
Due to smoothness, the $\phi(\lambda_Q, \lambda_W) = r$ iso-contour is $\partial A_r$,
the boundary of the sublevel set $A_r$ from Eq.~\eqref{eq:PhiSublevelSets}.
Since $\phi(\lambda_Q, \lambda_W)$ is strictly convex, so are the sublevel sets
$A_r$, and consequently the point $\tilde{\bm\lambda}(\eta)$ is unique for all $\eta$.
The fact that the line of efficiency $\eta$ is tangent to an iso-contour of
$\phi(\lambda_Q, \lambda_W)$ in $\bm\minLCurve(\eta)$ means that the ray vector
$(\eta, 1)$ of the line $\lambda_Q = \eta \lambda_W$ is orthogonal to the gradient of
$\phi(\lambda_Q, \lambda_W)$ in $\bm\minLCurve(\eta)$.
Thus
\begin{equation}
\label{eq:LambdaEta}
	\eta \frac{\partial\phi}{\partial\lambda_Q}(\minLCurve_Q(\eta), \minLCurve_W(\eta)) + \frac{\partial\phi}{\partial\lambda_W}(\minLCurve_Q(\eta), \minLCurve_W(\eta)) = 0 \,.
\end{equation}
This relation along with $\minLCurve_Q(\eta) = \eta \minLCurve_W(\eta)$
implicitly defines $\bm\minLCurve(\eta)$ as a function of $\eta$.
More precisely, we consider the function
\begin{equation}
	f(\eta, \lambda) := \eta \frac{\partial\phi}{\partial\lambda_Q}(\eta \lambda, \lambda) + \frac{\partial\phi}{\partial\lambda_W}(\eta \lambda, \lambda) \,.
\end{equation}
Assume that we have a particular solution $\lambda_*$ of~\eqref{eq:LambdaEta} for some given $\eta_*$, so that $f(\eta_*, \lambda_*) = 0$.
Due to the assumed positive definiteness of the Hessian matrix of $\phi(\lambda_Q, \lambda_W)$, we have
\begin{equation}
	\frac{\partial f}{\partial\lambda}(\eta_*, \lambda_*)
		= \left(\begin{matrix} \eta_* & 1 \end{matrix}\right)
			\left(\begin{matrix} \frac{\partial^2 \phi}{\partial\lambda_Q^2}(\eta_*, \lambda_*) & \frac{\partial^2 \phi}{\partial\lambda_Q \partial\lambda_W}(\eta_*, \lambda_*) \\ \frac{\partial^2 \phi}{\partial\lambda_W \partial\lambda_Q}(\eta_*, \lambda_*) & \frac{\partial^2 \phi}{\partial\lambda_Q^2}(\eta_*, \lambda_*) \end{matrix}\right)
			\left(\begin{matrix} \eta_* \\ 1 \end{matrix}\right)
		> 0 \,.
\end{equation}
By the implicit function theorem, there exists a function
$\eta \mapsto \tilde\lambda(\eta)$ on an open interval $I \subseteq \mathbb R$
with $\eta_* \in I$ and such that $f(\eta, \tilde\lambda(\eta)) = 0$ for all $\eta \in I$.
Moreover, this function is of the same differentiability class as $f(\eta, \lambda)$.
Put differently, the parameter function $\eta \mapsto \tilde\lambda(\eta)$ implicitly
defined by Eq.~\eqref{eq:LambdaEta} is well-defined and smooth in $I$.
As this holds everywhere in $U$, we conclude that
$\bm\minLCurve(\eta) = (\eta \tilde\lambda(\eta), \tilde\lambda(\eta))$
and thus $J(\eta) = -\phi(\minLCurve_Q(\eta), \minLCurve_W(\eta))$ is smooth.

\subsection*{Least likely efficiency}
The ``least likely'' efficiency, \emph{i.e.}\ the value $\hat\eta$ that
maximizes $J(\eta)$, is directly related to the global minimum of $\phi(\lambda_Q, \lambda_W)$.
Indeed, if $\phi(\lambda_Q, \lambda_W)$ assumes its minimal value at $(\hat\lambda_Q, \hat\lambda_W)$,
then $J(\eta)$ will become maximal for $\hat\eta = \hat\lambda_Q / \hat\lambda_W$ by Eq. (5). 
As observed in the main text and investigated in more detail in the next section of this
Supplemental Material, the global minimum of $\phi(\lambda_Q, \lambda_W)$ need not be unique
in general, so that $J(\eta)$ can have a degenerate maximum (``plateau'') as well.
In any case, the reversible efficiency $\effCarnot$ maximizes $J(\eta)$ if and only if
the global minimum of $\phi(\lambda_Q, \lambda_W)$ lies on the line $\lambda_Q = \effCarnot \lambda_W$.

As observed in Refs.~\cite{Verley:2014uce, Verley:2014ute}, the symmetry property Eq. (6)
provides a sufficient (but not necessary) condition for the least likeliness of the
reversible efficiency $\effCarnot$.
Indeed, if Eq.\ (6) holds then the iso-contour lines of $\phi(\lambda_Q, \lambda_W)$
are invariant under reflection through the point $(\lambda_Q^*/2, \lambda_W^*/2)$.
By convexity, $\phi(\lambda_Q, \lambda_W)$ must therefore attain its minimal value in the
reflection point, so that $\phi(\lambda_Q^*/2, \lambda_W^*/2) = \hat\phi$.
Hence $J(\eta)$ assumes the maximum possible value for the line with slope
$\hat\eta = \lambda_Q^* / \lambda_W^* = \effCarnot$.

\subsection*{Plateau}
We have illustrated in the main text that there could be cases where the global maximum of $J(\eta)$
is not unique, meaning that there may be an entire ``plateau region'' where $J(\eta)$
assumes its maximal value.
As stated in the main text,

\textit{The maximum of $J(\eta)$ is unique if and only if all minimizing points
$(\lambda_Q, \lambda_W)$ of $\phi$ lie on a line $\lambda_Q = \hat\eta \lambda_W$
through the origin with fixed slope $\hat\eta$.} 

Below, we elaborate on this feature.

Let us first assume that there exist
$\bm\lambda^{(1)}, \bm\lambda^{(2)} \in A_{\hat\phi}$ with $\bm\lambda^{(1)} \neq \bm\lambda^{(2)}$ and
$\lambda^{(1)}_Q / \lambda^{(1)}_W \neq \lambda^{(2)}_Q / \lambda^{(2)}_W$,
so that the minimizing points of $\phi(\lambda_Q, \lambda_W)$
do not lie on a single line through the origin.
(Recall that $A_{\hat\phi}$ denotes the set of all
$\bm\lambda = (\lambda_Q, \lambda_W)$ with
$\phi(\lambda_Q, \lambda_W) = \hat\phi$, \emph{c.f.}\ Eq.~\eqref{eq:PhiSublevelSets}).
The latter condition ensures that the slopes $\eta^{(i)} = \lambda^{(i)}_Q / \lambda^{(i)}_W$
of the lines connecting the origin with $\bm\lambda^{(1)}$ and $\bm\lambda^{(2)}$,
respectively, are different, meaning that $\eta^{(1)} \neq \eta^{(2)}$.
But since both lines cut through the global minimum, it follows that
$J(\eta^{(1)}) = J(\eta^{(2)}) = -\hat\phi$, establishing the degeneracy of $J(\eta)$.
Moreover, by convexity of $\phi(\lambda_Q, \lambda_W)$, all lines with slopes
$\eta$ between $\eta^{(1)}$ and $\eta^{(2)}$ will also cross the global minimum,
so that \emph{all plateau efficiencies are connected}.
In other words, there cannot be two separate plateaus in disjoint intervals of
the extended(!) real line $\mathbb R \cup \{\infty\}$.
(However, the plateau may be connected through the point $\eta = \pm\infty$,
corresponding to the line $\lambda_W = 0$.)
We remark that the emergence of plateaus need not necessarily be due to ``tight coupling''
as in the Brownian gyrator example presented in the main text (Fig.4 ).
The tight coupling case, where $\phi(\lambda_Q, \lambda_W) = \phi(\lambda_Q - \bar\eta \lambda_W)$
is a special case exhibiting a degenerate global minimum.

To show the converse direction, assume that there is a unique $\hat\eta$
such that all points $\bm{\hat\lambda} = (\hat\lambda_Q, \hat\lambda_W)$ with
$\phi(\hat\lambda_Q, \hat\lambda_W) = \hat\phi$ satisfy $\hat\lambda_Q / \hat\lambda_W = \hat\eta$.
Then all such points $\bm{\hat\lambda}$ lie on the line $\lambda_Q = \hat\eta \lambda_W$,
while for all $\eta \neq \hat\eta$ and all $\lambda \in \mathbb R$, $\phi(\eta\lambda, \lambda) > \hat\phi$.
Hence $J(\eta)$ has a unique maximum at $\hat\eta$.

\section*{Example 1: Isothermal work-to-work converter engine \cite{Gupta:2017sei}}
In this section, we provide details about the example of an isothermal work-to-work
converter by briefly summarizing the main results from Ref.~\cite{Gupta:2017sei}
that are relevant for our purposes.

The model consists of a Brownian particle of mass $m$ in a fluid environment at
temperature $T$ with instantaneous velocity $v(t)$.
By the fluctuation-dissipation theorem, the coupling to the heat bath
gives rise to a fluctuating force $\sqrt{2 k_{\mathrm{B} T \gamma} \gamma} \eta(t)$
as well as a frictional force $-\gamma v(t)$, where $\gamma$ is the friction
coefficient and $\eta(t)$ is a Gaussian white-noise process with
$\langle \eta(t) \rangle = 0$ and $\langle \eta(t) \eta(t') \rangle = \delta(t-t')$.
In addition, the particle is subject to two more fluctuating forces
$f_1(t)$ and $f_2(t)$ with Gaussian white-noise statistics, independent
of each other as well as of the thermal noise,
\emph{i.e.}\ $\langle f_i(t) f_j(t') \rangle = \delta_{ij} \bar f_i^2 \delta(t-t')$
and $\langle f_i(t) \eta(t') \rangle = 0$.
The resulting equation of motion thus reads
\begin{equation}
	m \dot{v}(t) = -\gamma v(t) + f_1(t) + f_2(t) + \sqrt{2 k_{\mathrm{B}} T \gamma} \eta(t) \,.
\end{equation}
The relative strength of the three fluctuating forces with respect to each other
is parameterized by the positive parameters $\theta$ and $\alpha$ such that
$\bar f_1^2 = 2 k_{\mathrm B} T \gamma \theta$ and $\bar f_2^2 = 2 k_{\mathrm B} T \gamma \theta \alpha^2$.
The force $f_1(t)$ and $f_2(t)$ are interpreted as a load and drive force, respectively.
The work done by them is given by
\begin{equation}
	W_i = \frac{1}{T} \int_0^\tau dt \; f_i(t) \, v(t) \,.
\end{equation}
Translated to the setting in the main text,
we thus identify $W_1$ with $W$ and $W_2$ with $Q_1$.
The resulting moment-generating function for $W_1$ and $W_2$
was found in Ref.~\cite{Gupta:2017sei} to satisfy the asymptotic relation
\begin{align}
\Psi_\tau(\lambda_1,\lambda_2) = \left\langle e^{-\lambda_1 W_1 - \lambda_2 W_2} \right\rangle_\tau
\sim g(\lambda_1,\lambda_2)\;e^{\tau\; \mu(\lambda_1,\lambda_2)}
\quad (\tau\to\infty),
\end{align}
where
\begin{align}
\label{d}
\mu(\lambda_1,\lambda_2)&=\frac{1}{2}\left[1- \nu(\lambda_1,\lambda_2)\right],\\
\nu(\lambda_1,\lambda_2)&=\left[1+4\theta\left\lbrace
\lambda_1(1-\lambda_1)+\alpha^2\lambda_2(1-\lambda_2)
-\alpha^2\theta (\lambda_1-\lambda_2)^2\right\rbrace\right]^{\frac{1}{2}},\\
g(\lambda_1,\lambda_2)&=\scalebox{0.9}{$ \sqrt{\frac{2\sqrt{-4 \alpha ^2 \theta ^2 (\lambda_1-\lambda_2)^2-4 \theta  \left(\alpha ^2 (\lambda_2-1) \lambda_2+(\lambda_1-1) \lambda_1\right)+1}}{\sqrt{-4 \alpha ^2 \theta ^2 (\lambda_1-\lambda_2)^2+4 \theta  \left(\alpha ^2 (-(\lambda_2-1)) \lambda_2-\lambda_1^2+\lambda_1\right)+1}+2 \theta  \left(-\lambda_1^2 \left(\alpha ^2 \theta +\theta +1\right)+\alpha ^2 (-\lambda_2) \left(\lambda_2 \left(\alpha ^2 \theta +\theta +1\right)-1\right)+\lambda_1\right)+1}}$} . 
\end{align}
From this, the scaled cumulant generating function $\phi(\lambda_1, \lambda_2) \equiv \phi(\lambda_Q, \lambda_W)$ can be extracted straightforwardly.

\section{Example 2: Brownian gyrator}
\label{sec:BronwianGyrator}

In this section, we give a detailed definition of the Brownian gyrator model
adapted from Ref.~\cite{Filliger:2007bgm} and provide the exact solution of
its scaled cumulant generating function $\phi(\lambda_Q, \lambda_W)$.

\begin{figure}
\centering
\includegraphics[scale=0.35]{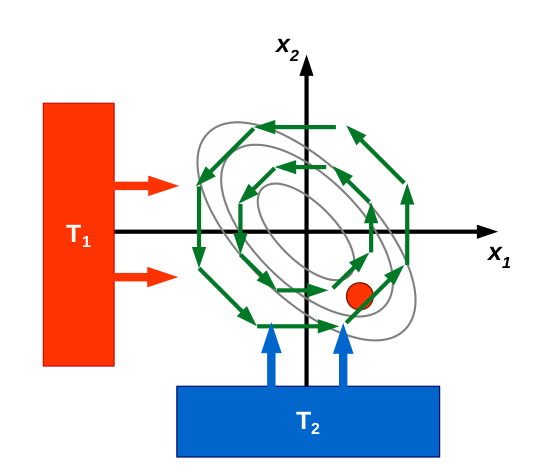}
\caption{Sketch of the Brownian gyrator: A Brownian particle in two dimensions, sits in a
potential $U$ (gray) and experiences an additional vortex force $\bm f_{\rm{ext}}$ (green).
It dissipates energy into a (fluid) medium via friction and is also subject to
thermally fluctuating forces from collisions with the fluid molecules.
The medium itself is in contact with two orthogonally irradiating reservoirs at
temperatures $T_1$ (red) and $T_2$ (blue), leading to different fluctuation intensities
in the two coordinate directions.}
\label{fig:Gyrator:Sketch}
\end{figure}

The model consists of a Brownian particle in two dimensions at
position $\textbf{x} = (x_1 \; x_ 2 )^T$, sketched in Fig.~\ref{fig:Gyrator:Sketch}.
The particle is immersed in a fluid environment and simultaneously
coupled to two (effective) heat baths at different temperatures $T_1 > T_2$
that only act in the $x_1$ and $x_2$ directions, respectively.
For example, the colder temperature $T_2$ may be the temperature of the surrounding
fluid, while there are additional fluctuations in the $x_1$ directions due to external fields
or an irradiating heat bath leading to a higher effective temperature $T_1$ \cite{Filliger:2007bgm},
see \cite{Argun:2017erm} for an experimental realization or \cite{Chiang:2017eab}
for an equivalent electric circuit system.

By the fluctuation-dissipation theorem, the coupling to the environments leads,
in both directions, to fluctuating forces $\sqrt{2 k_{\mathrm B} T_i \gamma_i} \eta_i(t)$
on the one hand and frictional forces $-\gamma_i \dot{x}_i$ on the other hand,
where $\gamma_i$ are the respective friction coefficients and $\xi_i(t)$ are
independent Gaussian white-noise processes with
$\langle \eta_i(t) \rangle = 0$ and $\langle \eta_i(t) \, \eta_j(t') \rangle = \delta_{ij} \, \delta(t - t')$.
The particle is confined by a parabolic potential $U(x)$ with stiffnesses $u_1$ and $u_2$
along its principal axes, which are tilted by an angle $\alpha$ with respect to the coordinate axes: 
\begin{align}
U(\textbf{x})&=\frac{1}{2}\;\textbf{x}^T\;\textbf{R}_\alpha^T\;\textbf{u}\;\textbf{R}_\alpha\;\textbf{x},&\textbf{R}_\alpha&=\left(\begin{array}{cc}
\cos \alpha&-\sin \alpha\\
\sin \alpha & \cos \alpha
\end{array}\right),&\textbf{u}&=\left(\begin{array}{cc}
u_1&0\\
0&u_2
\end{array}\right).
\end{align}
Due to the asymmetry of the thermal and restoring forces
(for $T_1 \neq T_2$, $u_1 \neq u_2$, and $\alpha \neq \pi n / 4$, $n \in \mathbb{Z}$),
the particle reaches a non-equilibrium steady state and rotates around the
origin on average \cite{Filliger:2007bgm}.
It thereby exerts a torque on the environment and can thus
work as a microscopic heat engine.
To quantify the work done, we generalize the model studied in \cite{Filliger:2007bgm}
by introducing an additional external force (see also \cite{Pietzonka:2018utp})
\begin{align}
\bm f_{\rm{ext}}(\textbf{x})&=-f_{\rm{ext}}\bm\epsilon\; \bm{x},&
\text{where }
\bm\epsilon=\left(\begin{array}{cc}
0&1\\-1&0
\end{array}\right)
\end{align}
is the two dimensional antisymmetric tensor. In the overdamped limit
the dynamics of the Brownian Gyrator is then described by the equations of motion
\begin{align}
\label{lang}
\dot{\bm x}(t)=-\textbf{A} \;\bm x(t)+\textbf{B}\;\bm\eta(t),
\end{align}
where
\begin{align}
\textbf{A}&=\left(\begin{array}{cc}
\frac{K_{11}}{\gamma_1}&\frac{K_{12}}{\gamma_1}\\
\frac{K_{21}}{\gamma_2}&\frac{K_{22}}{\gamma_2}
\end{array}\right),&\textbf{K}&=\textbf{R}_\alpha^T\;\textbf{u}\;\textbf{R}_\alpha+f_{\rm{ext}}\;\bm\epsilon,&\textbf{B}&=\left(\begin{array}{cc}
\sqrt{\frac{2k_BT_1}{\gamma_1}}&0\\0&\sqrt{\frac{2k_BT_2}{\gamma_2}}
\end{array}\right) .
\end{align}
For the range of parameter values where the matrix $\textbf{A}$ is positive definite,
the system reaches a steady state with probability distribution \cite{Argun:2017erm}
\begin{align}
p_{\mathrm{st}}(\bm x)=\frac{1}{2\pi\sqrt{\det \bm\Sigma (\infty)}}\;\exp(-\frac{1}{2}\;\bm{x}\;\bm\Sigma^{-1}(\infty)\;\bm{x}),
\end{align}
where $\bm\Sigma(\infty)$ is obtained as a solution of
\begin{align}
\textbf{A}\bm\Sigma(\infty) + \bm\Sigma(\infty)\textbf{A}^T &= 2\textbf{D},&\textbf{D}&=\frac{1}{2}\textbf{B}\textbf{B}^T.
\end{align}
The work done by the external load force $\bm f_{\mathrm{ext}}$
as well as the heat taken from the hot reservoir over a process of
time duration $\tau$ can be obtained using the standard definitions of
stochastic thermodynamics as
\begin{align}
W[\bm{x}(\cdot)]&=\sum_{i,j} \int_0^\tau Y^W_{ij}\;x_j\;dx_i,&Q_1[\bm{x}(\cdot)]&=\sum_{i,j} \int_0^\tau Y^{Q_1}_{ij}\;x_j\;dx_i
\end{align}
with
\begin{align}
\textbf{Y}^W&=-\rm{f}^{\rm{ext}}\;\bm\epsilon,&\textbf{Y}^{Q_1}&=\left(\begin{array}{cc}K_{11}&K_{12}\\0&0 
\end{array}\right) .
\end{align}
In Fig.~\ref{fig:Gyrator:Regimes}, we display the resulting average input
and output powers $\langle q_1 \rangle$, $\langle q_2 \rangle$, and $\langle w \rangle$ as a function of the load amplitude
$f_{\mathrm{ext}}$ for a certain choice of parameters.
It illustrates that the system indeed works as a heat engine for moderate
loads with $f_{\mathrm{ext}} = -1 \ldots 0$.

\begin{figure}
\centering
\includegraphics[scale=0.6]{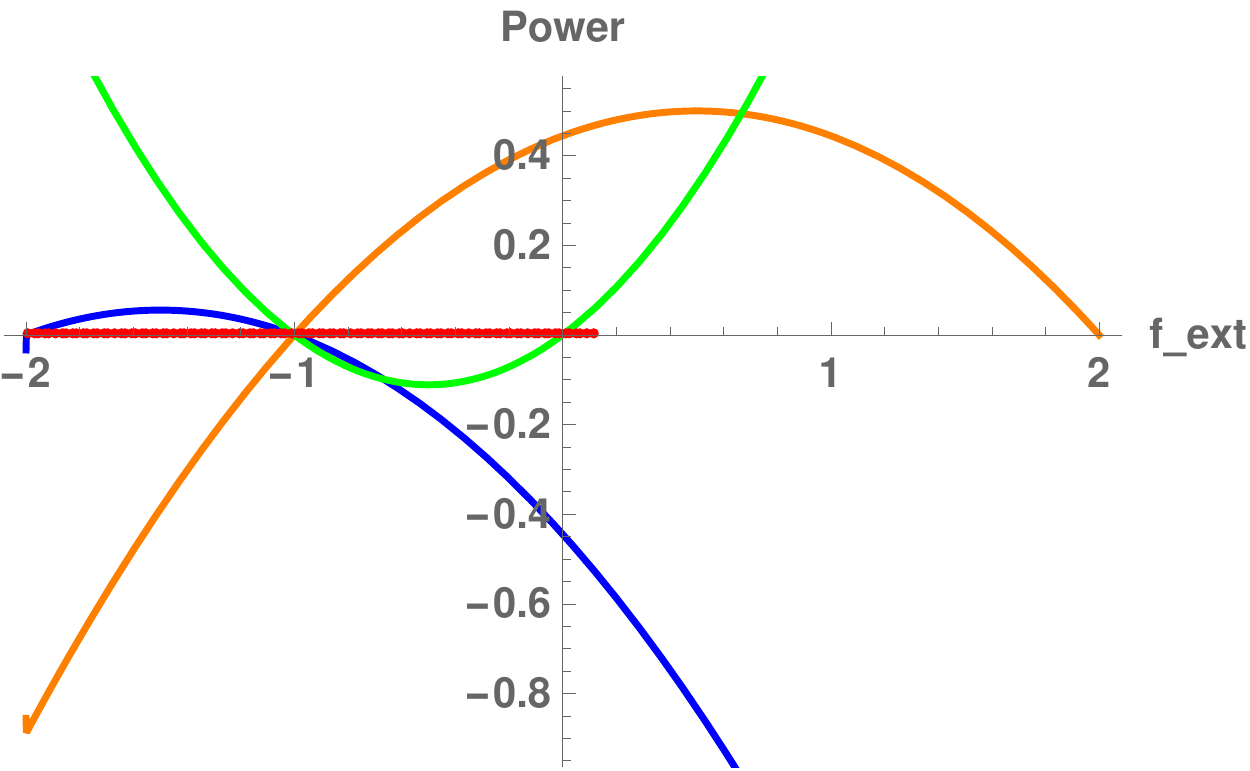}
\caption{
Average power supplied from the two reservoirs ($\langle q_1 \rangle$ orange, $\langle q_2 \rangle$ blue)
and by the external load force ($\langle w \rangle$ green).
The solid lines correspond to the analytical solution.
In the units chosen, the system reaches a steady state in the parameter range $f_{\rm{ext}} = - 2\text{ to } 2$. Further, it acts as a heat pump (refrigerator,
$\langle q_2 \rangle >0,\; \langle q_1 \rangle<0 \text{ and } \langle w \rangle>0$) between $f_{\rm{ext}} = - 2\text{ to } - 1$
and as a heat engine ($\langle q_2 \rangle <0,\; \langle q_1 \rangle>0 \text{ and } \langle w \rangle<0$) between
$f_{\rm{ext}} = - 1\text{ to } 0$. Beyond this point, we observe a trivial heat transfer from the hot to the
cold reservoir. For values of $f_{\rm{ext}}$ outside $f_{\rm{ext}} = - 2\text{ to } 2$, the external work trivially
heats both the hot and cold reservoirs, operating as a {\it dud engine}.
The range of $f_{\rm{ext}} = -2 \text{ to } 0.1$ that is marked red corresponds
to the case when the efficiency rate function $J(\eta)$ has an infinite plateau
as in the top panel of Fig.\ 4 from the main text. In the range of
$f_{\rm{ext}} = 0.1 \text{ to } 1$, $J(\eta)$ has a finite plateau as in the
lower panels of Fig.\ 4 from the main text. Parameter values: $k_B T_1 = 1$, $k_B T_2 = 1/3$, $u_1 = 5$, $u_2 = 1$, $\alpha =\pi/4$, $\gamma_1 =\gamma_2 = 1$.}
\label{fig:Gyrator:Regimes}
\end{figure}

Now using the path integral formalism, the moment generating function (MGF) of
$Q_1$ and $W$ at arbitrary times can be obtained as
\begin{align}
\label{ac}
\Psi_\tau(\lambda_Q, \lambda_W) =
\left\langle e^{-\lambda_Q Q_1-\lambda_W W}\right\rangle_\tau
=\int d\bm{x}_0 \; p_{\mathrm{st}}(\bm{x}_0)\int d\bm{x}_\tau \int_{\bm{x}_0}^{\bm{x}_\tau} D\bm{x}[(\cdot)]P[\bm{x}(\cdot)]e^{-\lambda_Q Q_1[\bm{x}(\cdot)]-\lambda_W W[\bm{x}(\cdot)]} \,,
\end{align}
where
\begin{align}
P[\bm{x}(\cdot)] \propto \exp\left(-\int_0^\tau dt\;\left(\; \dot{\bm{x}}(t)+\textbf{A}\bm{x}(t)\;\right)^T\frac{1}{2\textbf{D}}\left( \;\dot{\bm{x}}(t)+\textbf{A}\bm{x}(t)\;\right)\right)
\end{align}
denotes the Onsager-Machlup path weight \cite{Onsager:1953fai,Machlup:1953fai,Chernyak:2006pia}.
Since all the terms in the exponent of the RHS of Eq.\ \eqref{ac} are quadratic
in $x_1,\;x_2$ and their derivatives, we can write this as \cite{Manikandan:2017awd}
\begin{align}
\label{bcs}
\left\langle e^{-\lambda_Q Q_1[\bm{x}(\cdot)]-\lambda_W W[\bm{x}(\cdot)]}\right\rangle &=\int d\bm{x}_0\int d\bm{x}_\tau \int_{\bm{x}_0}^{\bm{x}_\tau} D\bm{x}[(\cdot)]\;\exp\left( \;\int_0^\tau\bm{x}(t)\;\hat{\textbf{O}}_{\lambda_Q,\lambda_W}\;\bm{x}(t)\;+\;\text{Boundary terms}\;\right)\\
\label{dets}
&=\sqrt{\frac{\det \;\hat{\textbf{O}}_{0,0}}{\det \;\hat{\textbf{O}}_{\lambda_Q,\lambda_W}}}.
\end{align}
Here the operator $\hat{\textbf{O}}$ is a matrix with differential operators as its entries \cite{Kirsten:2003fdc}, and the determinants that appear in Eq.\ \eqref{dets} are functional determinants. For our problem, it can be shown that the matrix $\hat{\textbf{O}}$ has the form
\begin{align}
\hat{\textbf{O}}=\left[\begin{array}{cc}
-a\;\frac{d^2}{dt^2}+b,&c\;\frac{d}{dt}+d\\
-c\;\frac{d}{dt}+d&-e\;\frac{d^2}{dt^2}+f
\end{array}\right],
\end{align}
where
\begin{align}
\begin{split}
a &= \frac{1}{4D_{11}},\\b &=\frac{1}{2}\left(\frac{A_{11}^2}{2 D_{11}} +\frac{A_{21}^2}{2 D_{22}} \right),\\c&=\frac{1}{2} \left(-\frac{A_{12}}{2D_{11}}+\frac{A_{21}}{2 D_{22}} \right)- \lambda_Q \frac{A_{12}}{2} + \lambda_W\;f_{\rm{ext}},\\d &=\frac{1}{2}\frac{A_{11}A_{12}}{2D_{11}} +\frac{1}{2}\frac{A_{21}A_{22}}{2D_{22}},\\ e &= 
 \frac{1}{4 D_{22}},\\ f &= 
 \frac{1}{2}\left(\frac{A_{12}^2}{2 D_{11}} +\frac{A_{22}^2}{2 D_{22}}\right).
 \end{split}
 \end{align}
Then the determinant ratio that appears in Eq.\ \eqref{dets} can be computed using a technique described in \cite{Kirsten:2003fdc} and recently used in \cite{Manikandan:2017awd}, which is based on the spectral-$\zeta$ functions of Sturm-Liouville type operators. Applying this method to the model at hand, it can be shown that this ratio can be obtained in terms of a characteristic polynomial function $F$ as
\begin{align}
\label{mgf}
\left\langle e^{-\lambda_Q Q_1[\bm{x}(\cdot)]-\lambda_W W[\bm{x}(\cdot)]}\right\rangle_\tau &=\sqrt{\frac{F(0,0)}{F(\lambda_Q,\lambda_W)}},  &F(\lambda_Q,\lambda_W) &\equiv \text{Det}\left[M+N H(\tau)\right].
\end{align}
Here $H$ is the matrix of independent and suitably normalized fundamental solutions $\bm x^1(t), \ldots, \bm x^4(t)$ of the homogeneous equation $\hat{\textbf{O}}\; \bm{x}=0 $,
\begin{align}
H(t)&=
\left[
\begin{array}{cccc}
 x_1^1(t) & x^2_1(t) & x_1^3(t) & x_1^4(t) \\
 x_2^1(t) & x_2^2(t) & x_2^3(t) & x_2^4(t)\\
 \dot{x}_1^1(t) & \dot{x}_1^2(t) & \dot{x}_1^3(t) & \dot{x}_1^4(t)\\
\dot{x}_2^1(t) & \dot{x}_2^2(t) & \dot{x}_2^3(t) & \dot{x}_2^4(t) \\
\end{array}
\right],& H(0) &=\textbf{I}_4,
\end{align}
and $M$ and $N$ contain information about the boundary conditions from Eq.\ \eqref{bcs}, for which we require
\begin{align}
M \left[\begin{array}{c}\bm{x}(0)\\\dot{\bm{x}}(0)\end{array}\right]&=0,&N \left[\begin{array}{c}\bm{x}(\tau)\\\dot{\bm{x}}(\tau)\end{array}\right]&=0.
\end{align}
A derivation of Eq.\ \eqref{mgf}, applicable to a class of driven Langevin systems with quadratic actions, is given in \cite{Manikandan:2017awd}.
We stress that the expression given in Eq.\ \eqref{mgf} is valid within the domain $C_{\lambda_Q,\;\lambda_W}$ for which the operator $\hat{\textbf{O}}$ doesn't have negative eigenvalues. The MGF is not convergent outside this domain.\par
For the Brownian gyrator problem, we obtain the four independent solutions as
\begin{align}
x_1^i(t)&=\exp(\;\pm\; t\;\frac{\sqrt{\pm\frac{\sqrt{a^2 f^2-2 a b e f-2 a c^2 f+4 a d^2 e+b^2 e^2-2 b c^2 e+c^4}}{a e}+\frac{b}{a}-\frac{c^2}{a e}+\frac{f}{e}}}{\sqrt{2}}\;) ,\\
x_2^i(t)&=\frac{x_1^i(t) \left(-\left(c^2 d-a\; d\; f\right)\right)+c \left(a\; f-c^2\right) x_1^{i\;\prime}(t)-a\; c\; e\; x_1^{i\;\prime\prime\prime}(t)-a\; d\; e \;x_1^{i\;\prime\prime}(t)(t)}{b c^2-a d^2} .
\end{align}
The matrices $M$ and $N$ are given by
\begin{align}
M&=\scalebox{0.8}{$\left(
\begin{array}{cccc}
 -\frac{2 D_{11} \lambda_Q A_{11}+A_{11}-2 D_{11} \Sigma_{11}}{4 D_{11}} & -\frac{2 D_{11} \lambda_Q A_{12}+A_{12}-2 D_{11} \text{f}_\text{ext} \lambda_W-2 D_{11} \Sigma_{12}}{4 D_{11}} & -\frac{1}{4 D_{11}} & 0 \\
 -\frac{A_{21}-2 D_{22} (\Sigma_{21}-\text{f}_\text{ext} \lambda_W)}{4 D_{22}} & -\frac{A_{22}-2 D_{22} \Sigma_{22}}{4 D_{22}} & 0 & -\frac{1}{4 D_{22}} \\
 0 & 0 & 0 & 0 \\
 0 & 0 & 0 & 0 \\
\end{array}
\right) ,$}
\\
N&=\scalebox{0.8}{$\left(
\begin{array}{cccc}
 0 & 0 & 0 & 0 \\
 0 & 0 & 0 & 0 \\
 \frac{2 D_{11} \lambda_Q A_{11}+A_{11}}{4 D_{11}} & \frac{2 D_{11} \lambda_Q A_{12}+A_{12}-2 D_{11} \text{f}_\text{ext} \lambda_W}{4 D_{11}} & \frac{1}{4 D_{11}} & 0 \\
 \frac{A_{21}+2 D_{22} \text{f}_\text{ext} \lambda_W}{4 D_{22}} & \frac{A_{22}}{4 D_{22}} & 0 & \frac{1}{4 D_{22}} \\
\end{array}
\right)$}.
\end{align}
Using these, the MGF can be computed exactly using Eq.\ \eqref{mgf}. Notice that the solution is valid for arbitrary time $\tau$. Various interesting finite time aspects of this solution will be discussed in a future publication. Here we focus on the large time limit, where the leading order form of the MGF is given by
\begin{align}
\label{tim}
\begin{split}
\Psi_\tau(\lambda_Q,\lambda_W)
&\sim g(\lambda_Q,\lambda_W)\;e^{\tau\;\phi(\lambda_Q,\lambda_W)}.
\end{split}
\end{align}
Using the exact solution obtained from Eq.\ \eqref{mgf}, the large time functional form given above can be obtained by performing an asymptotic expansion using \textit{Mathematica}. We provide here the exact functional forms for the completion of the discussion:
\begin{align}
\scalebox{0.8}{$\phi(\lambda_Q,\lambda_W)=\frac{1}{8} \sqrt{\frac{D_{22}^2 \left(16 D_{11}^2 \left(2 a e \sqrt{\frac{b f-d^2}{a e}}+a f+b e\right)-A_{12}^2\right)+2 A_{12} A_{21} D_{11} D_{22}-A_{21}^2 D_{11}^2}{a D_{11}^2 D_{22}^2 e}}-\sqrt{-\frac{\left(-\frac{A_{12}}{D_{11}}-2 A_{12} \lambda_Q+\frac{A_{21}}{D_{22}}+4 \text{f}_{\text{ext}} \lambda_W\right)^2}{64 a e}+\frac{1}{4} \sqrt{\frac{4 b f}{a e}-\frac{4 d^2}{a e}}+\frac{b}{4 a}+\frac{f}{4 e}}$}.
\end{align}
In terms of the function $\Gamma(\lambda_Q,\lambda_W)$ defined as
\begin{align}
\begin{split}
&\scalebox{0.75}
{$\Gamma(\lambda_Q,\lambda_W)=$}\\&\scalebox{0.75}{$\frac{-ae}{32 a e^2 \left(a d^2-b c^2\right) \sqrt{\frac{b f-d^2}{a e}} \left(2 \left(a \left(2 e \sqrt{\frac{b f-d^2}{a e}}+f\right)+b e-c^2\right)\right)}\bigg[\sqrt{-\frac{-2 a e \sqrt{\frac{b f-d^2}{a e}}-a f-b e+c^2}{a e}} \left(4 e \left(a \left(c m s \sqrt{\frac{b f-d^2}{a e}}+d \left(o s \sqrt{\frac{b f-d^2}{a e}}+l m-n p\right)\right)+c o (b l-d p)\right)\right)+$}\\
&\scalebox{0.65}{$4 \left(b \left(c \left(e \left(o s \sqrt{\frac{b f-d^2}{a e}}+l m-n p\right)+f o s\right)+c^2 n s+d e l o\right)+c^2 (-e) o p \sqrt{\frac{b f-d^2}{a e}}+a \left(c e (l m-n p) \sqrt{\frac{b f-d^2}{a e}}+d \left(e (l o+m s) \sqrt{\frac{b f-d^2}{a e}}-d n s+f m s\right)\right)-c^2 d m s-c d^2 o s-d^2 e o p\right)$}\bigg]\\
&\bigg[\scalebox{0.75}{$\sqrt{\frac{a \left(2 e \sqrt{\frac{b f-d^2}{a e}}+f\right)+b e-c^2}{a e}} \left(-4 e \left(a \left(c i y \sqrt{\frac{b f-d^2}{a e}}+d \left(v y \sqrt{\frac{b f-d^2}{a e}}+i x-u w\right)\right)+c v (b x-d w)\right)\right)+$}\\
&\scalebox{0.65}{$4 \left(b \left(c \left(e \left(v y \sqrt{\frac{b f-d^2}{a e}}+i x-u w\right)+f v y\right)+c^2 u y+d e v x\right)+c^2 (-e) v w \sqrt{\frac{b f-d^2}{a e}}+a \left(c e (i x-u w) \sqrt{\frac{b f-d^2}{a e}}+d \left(e (i y+v x) \sqrt{\frac{b f-d^2}{a e}}-d u y+f i y\right)\right)-c^2 d i y-c d^2 v y-d^2 e v w\right)$}\bigg],\\\\
&\text{where }\scalebox{0.8}{$\left(
\begin{array}{ccc}
 m & n & o \\
 p & l & s \\
\end{array}
\right)=\left(
\begin{array}{ccc}
 N_{31} & N_{32} & N_{33} \\
 N_{41} & N_{42} & N_{44} \\
\end{array}
\right), \hspace*{5mm}\left(
\begin{array}{ccc}
 i & u & v \\
 w & x & y \\
\end{array}
\right)=\left(
\begin{array}{ccc}
 M_{11} & M_{12} & M_{13} \\
 M_{21} & M_{22} & M_{24} \\
\end{array}
\right)$},
\end{split}
\end{align}
we obtain
\begin{align}
\scalebox{1}{$g(\lambda_Q,\lambda_W)=\sqrt{\frac{\Gamma(0,0)}{\Gamma(\lambda_Q,\lambda_W)}}$}.
\end{align}

\end{widetext}

\end{document}